\def\lapprox{\lesssim}
\def\gapprox{\gtrsim}
\def\etal{{et~al. }}
\def\asec{\arcsec}
\def\amin{\arcmin}
\def\han2{{H$\alpha$+[NII]}}

\documentstyle[12pt,aasms]{article}

\journalid{337}{15 January 1989}
\articleid{11}{14}

\begin{document}

\title{Large-Scale Outflows in Edge-on Seyfert Galaxies. \\
       I. Optical Emission-line Imaging and Optical Spectroscopy}

\author{Edward J. M. Colbert\altaffilmark{1,2},
        Stefi A. Baum\altaffilmark{1},
        Jack F. Gallimore\altaffilmark{1,2},
        Christopher P. O'Dea\altaffilmark{1},
        Matthew D. Lehnert\altaffilmark{3},
        Zlatan I. Tsvetanov\altaffilmark{4}
        John S. Mulchaey\altaffilmark{5},
        Saul Caganoff\altaffilmark{6},
        }

\altaffiltext{1}{ Space Telescope Science Institute, 3700 San Martin Drive,
                  Baltimore, MD  21218  }
\altaffiltext{2}{ Department of Astronomy, University of Maryland,
                  College Park, MD  20742}
\altaffiltext{3}{ Lawrence Livermore National Laboratories,
                  Institute of Geophys. \& Plan. Phys.,
                  7000 East Avenue, P.O. Box 808, L$-$413, Livermore, CA
94550}
\altaffiltext{4}{ Center for Astrophysical Sciences,
                  Department of Physics and Astronomy,
                  The Johns Hopkins University, 3701 San Martin Dr.,
                  Baltimore, MD 21218}
\altaffiltext{5}{ Carnegie Observatories, 813 Santa Barbara Street,
                  Pasadena, CA 91101}
\altaffiltext{6}{ ARC Systems Pty. Ltd., 55 Sussex St., Sydney, 2000,
Australia}

\begin{abstract}
We have launched a search for large-scale ($\gapprox$1 kpc) minor-axis outflows
in edge-on Seyfert galaxies
in order to assess their frequency of occurrence and study their properties.
Here we present optical continuum and \han2 line images and/or minor-axis
long-slit spectra of 22 edge-on Seyfert galaxies.
Six of these galaxies show at least one of the following:
(i) bi-symmetric H$\alpha$ halos extending along the minor axis,
(ii) bright emission-line complexes at distances $\gapprox$4 kpc
(in projection) out of the disk, and
(iii) double-peaked emission-line profiles from the gas along the minor-axis,
suggesting that a wind-blown bubble is present.
Our results indicate that $\gapprox$${{1}\over{4}}$ of Seyferts have
good evidence for minor-axis galactic outflows.

Kinetic luminosities of the galactic outflows in our sample Seyferts are
$\sim$10$^{40}$$-$10$^{42}$ erg~s$^{-1}$, assuming all of the observed
minor-axis emission is produced by the outflow.  These values are,
in general, $\sim$0.1 as large as those for well-studied cases of
superwinds in starburst galaxies (Heckman, Armus \& Miley 1990).
However, far-infrared luminosities of our sample Seyferts are also
$\sim$0.1 as large.  Both starburst-driven superwinds and wide-angled
outflows from the active galactic nucleus are possible explanations
for the observed large-scale outflows.

\end{abstract}


\section{Introduction}

Ionized gas is known to extend (up to $\approx$1 kpc) out of the plane in
normal spiral galaxies (e.g. Dettmar 1992).
The origin and maintenance of this component of the interstellar medium in
normal galaxies is not completely understood, but most theories support
the notion that it is energized by outflows
from the galaxy disk (cf. Rand 1995).

In powerful infrared (IR) galaxies, extraplanar ionized gas is much more
luminous and is detectable much further (up to tens of kpc, Heckman,
Armus \& Miley 1990) out into the halo.
The origin of these more extended structures is also explained by outflows
from the galaxy.  Nuclear starbursts in these galaxies produce stellar winds
and supernovae which rapidly heat the gas in the nuclear region.  This
high-pressure gas then expands rapidly and blows a wind out of the nuclear
region,
along the rotation axis of the disk, where the pressure gradient is lowest.
This `superwind' interacts with clouds of dense gas in the galaxy and its halo,
producing optical emission-line filaments
(see, e.g., Heckman, Armus \& Miley 1990).
Material swept up by the wind accumulates behind the
shock front, forming a large shell (or `bubble') along the minor axis.
Line emission which is observed from the shell should then
have two velocity components, one from the front surface of the shell and one
from the rear surface.
Such double-peaked emission-line profiles
are indeed commonly found in minor axis
spectra of superwind galaxies (cf. Heckman, Lehnert \& Armus 1993),
corroborating that wind-blown bubbles are present.

Superwinds also manifest themselves at radio and X-ray wavelengths.  The
radio halos in edge-on starburst galaxies produce synchrotron
emission from relativistic electrons in the wind.
Thermal X-rays are emitted by hot, expanding gas in
the bubble and by low-density halo clouds which have been shocked by the wind.
A well studied example of an extended halo is that in the edge-on starburst
galaxy M82  (radio: e.g., Seaquist \& Odegard 1991; X-ray: Bregman 1994).
More examples are discussed in a
recent review of observational and theoretical aspects of superwinds in
starburst galaxies by Heckman, Lehnert \& Armus (1993).

Approximately 1\% of spiral galaxies have an active (Seyfert) nucleus,
which is commonly believed to be an accretion-powered supermassive black
hole, i.e., a scaled-down version of a quasar
(see, e.g., Terlevich 1992 for an alternate view).
The emission from the active galactic nucleus (AGN) dominates the luminosity
at most wavelengths, so
most studies of Seyferts have focused on observing the properties of the
AGN itself and the environment it creates in and immediately
around the nuclear region (see, e.g., Antonucci 1993).  There is good evidence
that nuclear outflows exist on pc scales in some Seyferts (cf. Wilson 1993),
but it is not clear that these small-scale ($\lapprox$1 kpc)
outflows are connected with large-scale galactic outflows.

Observations of the large-scale ($\gapprox$1 kpc) emission in Seyferts do
show that, in some cases, galactic-scale minor-axis outflows are present
(e.g., Hummel, van Gorkom \& Kotanyi 1983, Hamilton \& Keel 1987, Corbin
\etal 1988, Wehrle \& Morris 1988).
In a sample of Seyfert galaxies selected for extended optical or radio
emission, Baum \etal (1993) found kpc-scale radio emission extending
preferentially along the minor axes in seven out of ten cases and suggested
that the emission comes from galactic winds blowing out of the galaxy disks.
A particularly good example of a galactic outflow
can be found in the edge-on Seyfert galaxy NGC~5506.
Diffuse radio emission is found out to $\sim$300~pc from the nucleus in the
direction of the minor axis
(Wehrle \& Morris 1987) and double-peaked emission-line profiles
are found in minor axis spectra from regions $\sim$500 pc above and below the
disk (Wilson, Baldwin \& Ulvestad 1985).  The question of what powers
galactic outflows in Seyferts remains open, but viable sources of kinetic
energy are nuclear starbursts and the active nucleus.

Why study galactic outflows in Seyferts?  So far, only anecdotal evidence
has been presented for them
and they have not been systematically studied.  One would
like to know
how galactic outflows in Seyferts are different from those in more powerful
AGNs (e.g. radio galaxies, BALQSOs) and also from those in starburst galaxies.
{}From a broader standpoint, one would like to know what influence the
outflows have on the AGN and its host galaxy.  Galactic outflows drive large
amounts of gas (and metals) out of the nuclear region, whereas the AGN needs
a constant supply of gas flowing inward.  Thus, galactic outflows in Seyferts
may regulate the power output and lifetime of the active nucleus.
If the outflows are driven by starbursts, then what role, if any, do these
nuclear starbursts play in the formation and evolution of the active nucleus?
Many
workers have proposed scenarios in which Seyferts and starbursts are connected
in an evolutionary sense (e.g. Scoville 1992), but the issue remains open.
The outflows may also have a strong influence on the physical state of the
intergalactic medium.  Plasma which is blown out of the galaxy will increase
the density of the intergalactic medium, heat it, and enhance its metallicity.

In order to address these questions,
we have begun a research program to systematically
study a complete sample of edge-on Seyfert galaxies.
We shall search for observational evidence of galactic outflows,
determine their physical properties and investigate possible energy sources
for driving the outflows.
In paper~I, we present optical emission-line images and/or minor-axis
spectra of the ionized gas in 22 edge-on Seyfert galaxies.
Our future work will consist of analyzing the large-scale radio and X-ray
emission from these outflows, using the same sample of edge-on Seyferts.

In section 2, we introduce a statistical sample of Seyfert galaxies which
we shall use for continuing studies.  Observations and data reduction
techniques are described in section 3 and images and spectra are
presented for individual objects in section 4.
In section 5, we discuss the frequency of occurrence of
large-scale outflows in Seyferts and discuss possible interpretations of
energy sources driving the outflows.  A Hubble constant of
75 km~s$^{-1}$~Mpc$^{-1}$ is assumed throughout this paper.

\section{Definitions of Samples}

Extended optical emission-line gas is difficult to observe in face-on
disk galaxies, due to confusion with emission from the disk.  We have
therefore chosen to study only edge-on systems.

We have defined a `complete,' distance-limited statistical
sample of edge-on Seyferts which we shall use in continuing studies.
First, we selected Seyferts (LINERs were omitted) from
an electronic version of the Seyfert catalog by Huchra (1993).
This list was cross-referenced by position with an electronic version
(vers. 3.9) of
the {\it Third Reference Catalog of Bright Galaxies}
(de~Vaucouleurs \etal 1991;RC3).  Using recessional
velocities ($cz$; 21~cm preferred) and axial ratios ($R_{25}$; major to minor)
from RC3, we then selected objects by restricting
$cz \leq$ 5500 km~s$^{-1}$ and $\log R_{25} \geq$ 0.4
($R_{25} \geq$ 2.5).
Using the same method, we made another list starting from an electronic
version of Veron-Cetty \& Veron's (1991) Seyfert catalog.
The two lists were then combined.
A few additional Seyferts which satisfied the restrictions listed
above were selected by hand
from Hewitt \& Burbidge (1991) and from the literature.
Our complete sample consists of 22 objects.  Morphological types, positions,
axial ratios, recessional velocities
and Seyfert types for each galaxy are listed in Table 1.

We have obtained \han2 images and/or long-slit spectra (near H$\alpha$)
for 14 of these 22
objects.  No bias was used to select these objects for observation.
Henceforth, we will refer to this sub-sample (marked by a dagger [\dag] in
Table 1) as our `representative' sample
and shall use it for computing statistics in section 5.

We have also obtained images and/or spectra for eight additional
edge-on Seyferts which have
$cz >$ 5500 km~s$^{-1}$, $\log R_{25} <$ 0.4, or both.
Morphological types,
positions, axial ratios, recessional velocities and
Seyfert types for these galaxies are listed in
Table 2.

Hereafter we shall refer to the sample of all objects observed as
our `extended' sample.  These 22 objects are the 14 objects marked by a
dagger in Table 1 plus the eight objects listed in Table 2.  We shall use
this extended sample for computing statistics as well (section 5).

\section{Observations and Data Reduction}

A log of new imaging and spectroscopic observations using the
KPNO\footnotemark[1]\
\footnotetext[1]{
  Kitt Peak National Observatory of the National Optical Astronomy
  Observatories (NOAO), operated by AURA, Inc., under contract with the
  National Science Foundation.
}
2.1~m telescope is given in Table 3.
These data were supplemented by additional \han2 images,
taken at the 2.1~m telescope at KPNO, the
1.5~m telescope at the Cerro Telolo Inter-American Observatory (CTIO),
and the 2.2~m telescope at the European Southern Observatory (ESO).
Information about the observations for these supplemental images can be found
in Mulchaey, Wilson \& Tsvetanov (1996) and
Tsvetanov, Fosbury \& Tadhunter (1995).

Reduction of the data was performed using the CCDPROC analysis programs in
IRAF\footnotemark[2]\
\footnotetext[2]{
  IRAF is the imaging analysis software developed by NOAO.
  }.
The raw CCD images were processed in the normal fashion by subtracting bias
frames and dividing by normalized images of the dome or blank sky.
Atmospheric extinction corrections were performed (for all of the data)
using extinction curves appropriate for each observing site.

During the imaging runs at the KPNO 2.1~m (August 1992 and February 1993),
two
exposures of 1800 s were taken through narrow-band
(full width at half maximum [FWHM] $\sim$ 75\AA) filters centered near
H$\alpha$.
To produce an image of only the \han2 line emission,
an R-band image (exposure 300 s) was scaled and subtracted from the narrow-band
image.
Observations of spectrophotometric standard stars
were used to flux calibrate the \han2 images.  The limiting surface brightness
levels of the \han2 images are
$\sim$1 $\times$ 10$^{-16}$ erg~s$^{-1}$~cm$^{-2}$~arcsec$^{-2}$.

The long-slit spectra were taken
in January and June 1992 at the KPNO 2.1~m telescope.
Two exposures of 1800~s were taken with the slit oriented
parallel to the galaxy's minor axis and passing
through its nucleus.  One exposure of 1800 s was taken along
the major axis. Frequent observations of a quartz lamp were
used to remove the high spatial frequency structure.  HeNeAr spectra
were used for wavelength calibration and observations of
spectrophotometric standard stars
were used for absolute flux calibration.  The width of the slit was
1.75\asec, except for Ark~79, NGC~931, NGC~5506 and IC~1368, for which
it was 2.0\asec.  Limiting surface brightness levels at H$\alpha$ are
a factor of several better that those for the imaging data.  The spectral
coverage included H$\alpha$, [N~II] $\lambda\lambda$ 6548,~6583 and
[S~II] $\lambda\lambda$ 6716,~6731.

Continuum-free \han2 images from
Mulchaey, Wilson \& Tsvetanov (1996) and Tsvetanov, Fosbury \& Tadhunter (1995)
were made by combining narrow-band (FWHM $\sim$50$-$100\AA) images
centered on H$\alpha$ with narrow-band images of the nearby continuum.
The images were flux calibrated from observations of
spectrophotometric standard stars.
Exposure times were typically $\sim$1000$-$2000~s and the limiting
surface brightness levels were $\sim$2 $\times$ 10$^{-16}$
erg~s$^{-1}$~cm$^{-2}$~arcsec$^{-2}$ (except for the \han2 image of NGC~5506,
for which the exposure time was much shorter and the limiting
surface brightness is a factor of $\sim$10 higher).

\section{Results}

Descriptions of the observational data are given in Table 4.
For each of the 22 objects observed, we list the type of data
taken and the corresponding figure number for the images and/or spectra.
We also list the total \han2 luminosity from the galaxy
for all objects for which
we have \han2 images, and
the total \han2 luminosity from the minor-axis emission-line gas
for galaxies in which emission-line regions (ELRs)
 are evident along the minor axis.
The total \han2 luminosity from the galaxy was calculated by summing the flux
inside rectangular
regions enclosing all of the emission-line nebulae from the galaxy.
The \han2 luminosities of the minor-axis ELRs
were calculated by summing flux inside rectangular
regions around the individual ELRs mentioned in sections 4.1 and 4.2.

In the following subsections, we present and discuss our new images and
spectra and also note whether other data which has been published
is suitable for searching for evidence for minor-axis outflows. Seyferts which
are good candidates for having such outflows are identified and statistical
results for our samples are discussed in section 5.

\subsection{Individual Objects: Complete Sample}

For each of the galaxies in our complete sample, we searched the literature
for published data which could be used to look for signatures of minor-axis
outflows.  For many of the objects
(IC~1657, UM~319, Mrk~993, Mrk~577, Ark~79, NGC~1320, MCG~$-$2-27-9, NGC~4602,
ESO~103-G35, NGC~6810, IC~1417, NGC~7410 and NGC~7590),
the published data was not suitable.
Exceptions are noted in the following subsections.

\subsubsection{Mrk~993}

Our images for Mrk~993 are shown in Figure 1a.  Many ELRs are scattered
throughout the galactic disk, but no minor-axis emission is evident in
our \han2 image.
We did not find double-peaked line profiles or
any evidence for ELRs extending out of the disk  in our minor axis spectra.

\subsubsection{Ark~79}

No extraplanar emission is noticeable in our \han2 image
(Figure 1b).  The spectrum from the position 3.1\asec\ ($\sim$1 kpc)
south (along the minor axis) of the nucleus may have double-peaked
H$\alpha$, [NII] and [S~II] emission line profiles
(consistent with what is expected from the shell of an wind-blown bubble),
but the possible substructure in the profiles is nearly the
same level of magnitude as the noise.  For this reason, we have classified
these features as only suggestive of a minor axis outflow.
Spectra with higher signal-to-noise (S/N) ratios would be
useful for determining if these components are double peaked
emission lines from a wind-blown bubble.

\subsubsection{NGC~931 (Mrk~1040)}

Our R-band image (Figure 1c) shows the presence of a companion
galaxy $\sim$20\asec\ to the north along the minor axis.
Apart from emission from this companion galaxy, we find no extended
\han2 emission in our long-slit spectra along the minor axis.
Fabry-Perot studies (e.g. Amram \etal 1992) show that the kinematics of the
emission-line gas are quite complex, but there is no evidence for an outflow
from the nucleus.

\subsubsection{NGC 1320 (Mrk 607)}

No minor axis emission is noticeable in our \han2 image (Figure 1d).
This galaxy forms a pair with NGC~1321, 
which is positioned $\sim$1.7\amin\ to the north.

\subsubsection{NGC 1386}

NGC~1386 is located in the Fornax cluster, so we have
used the distance to that cluster (20 Mpc) in our calculations.
In the \han2 image (Figure 1e), there
is an ELR with luminosity 3.1 $\times$ 10$^{38}$ erg~s$^{-1}$  to the
northwest, at a distance of 41\asec\ (4.0 kpc) from the nucleus.  The
continuum emission from the disk
extends out to this position, so the ELR could lie in the the plane of
the galaxy disk.  Alternatively, the emission could be from gas in
a halo cloud which has been ionized by a wind.
Weaver, Wilson \& Baldwin (1991) have used long-slit spectroscopy
to study the kinematics of the extranuclear gas within $\sim$12\asec\
($\sim$1 kpc).
The line profiles and line ratios of the ELRs surrounding the nucleus
suggest that an outflow from the nucleus may be occurring along the
{\it major} axis.  Ulvestad \& Wilson (1984) found that the morphology of
the nuclear radio continuum emission is slightly extended
$\sim$ 400 pc toward the southwest along position angle (P.A.) $-$125\deg.
Although there is good
evidence for a nuclear outflow in NGC~1386, it does not appear to be directed
along the minor axis.  Except for the northwest ELR, all of the line emission
originates from $\lapprox$1 kpc.  More solid evidence is needed to conclude
that a galactic scale outflow (i.e. one blowing out of the galactic disk)
is present in this galaxy.

\subsubsection{NGC 2992}

A dust lane separates the line emission which immediately surrounds the
nucleus (see Figure 1f) from the bright
(L$_{H\alpha+[NII]} =$ 3.6 $\times$ 10$^{40}$ erg~s$^{-1}$)
ELR which extends from $\sim$5$-$15\asec\ (0.6$-$2.2 kpc)
to the northwest, in the direction of the minor axis.
A number
of ionized filaments of luminosity $\sim$10$^{38}$ erg~s$^{-1}$ are also
present beyond the northwest ELR, at distances $\le$30\asec\ (4.5 kpc).
Such bright filamentary structures are commonly found in
luminous IR galaxies
with starburst-driven superwinds (e.g. Heckman, Armus \& Miley 1990).
Tsvetanov, Dopita \& Allen (1995) find evidence for outflowing gas with
velocities $\sim$200 km~s$^{-1}$ on both sides of the galaxy disk.
Radio continuum maps of the nuclear region of NGC~2992 show emission
extending out to 2\asec\ (0.3 kpc) along the minor axis which
has the structure of ``striking pair of loops'' (Wehrle \& Morris 1988).
These authors interpret these loops as limb-brightened bubbles or magnetic
arches.  The bright minor-axis optical ELRs and `Figure-8' radio morphology
suggest that a powerful galactic outflow is occurring in NGC~2992.

\subsubsection{MCG~$-$2-27-9 }

Our \han2 image (Figure 1g) shows emission from the disk but no
extended emission along the minor axis.

\subsubsection{NGC 4235 (IC 3098)}

No
extended minor-axis
emission is apparent in our \han2 image (Figure 1h) of NGC~4235.
ELRs are noticeable
extending from the nucleus out along the {\it major} axis,
especially to the northeast (see also Pogge 1989).
We did not find double-peaked line profiles or
any evidence for ELRs extending out of the disk  in our minor axis spectra.
Radio continuum maps of the nuclear region (Ulvestad \& Wilson 1989) are
unresolved or very slightly resolved.  No extended emission was noticed by
Hummel, Beck \& Dettmar (1991) in their large-scale radio map.

\subsubsection{NGC~4388}

We did not obtain any new images or spectra of NGC~4388 since there is
already good evidence for a galactic outflow in this galaxy.
Corbin, Baldwin \& Wilson (1988) and Pogge (1988) have shown that conical
structures of ionized gas extend outward from the nucleus, both above and
below the disk.
These authors argue that some of the line-emitting gas may have been
ejected from the nucleus.
Radio maps of this galaxy
(Hummel \etal 1983; Stone, Wilson \& Ward 1988)
reveal diffuse emission extending outward from the nucleus,
perpendicular to the galactic disk.

\subsubsection{NGC 4602}

Much of the emission in our \han2 image (Figure 1i)
comes from a ring structure in the disk (Buta \& de~Vaucouleurs 1983), but
no emission is evident along the minor axis.

\subsubsection{NGC~4945}

This Seyfert galaxy also houses a strong nuclear starburst.
The presence of a superwind in this this galaxy has been discussed in detail
by Heckman, Armus \& Miley (1990), who argue that the wind is driven by the
starburst.  Harnett \etal (1989) present images of a radio halo in this
galaxy, presumably produced by the wind.

\subsubsection{IC 4329A}

A bi-symmetric halo of
emission-line gas can be seen in our \han2 image (Figure 1j)
extending along the minor axis, $\sim$10\asec\ (3 kpc) on both sides of the
nucleus.
The H$\alpha$+[NII] luminosity of the extended emission on either
side is $\sim$2.5 $\times$ 10$^{39}$ erg~s$^{-1}$.
Unger \etal (1987) show that radio emission extends $\sim$6\asec\ west from
the nucleus and suggest that it is from material lying out of the plane.
The H$\alpha$ halo we observe here is probably produced by an outflow from
the nuclear region and we consider
IC~4329A a good candidate for a large-scale galactic outflow.

\subsubsection{NGC 5506}

Emission-line nebulae extend along the minor axis, $\sim$5\asec\
(0.6 kpc) north and south of the nuclear region (see Figure 1k).
Wilson, Baldwin \& Ulvestad (1985) first noted double-peaked
[O~III] $\lambda$5007 and H$\beta$ emission lines in
spectra of the ELRs $\gapprox$5\asec\ north and south of the nuclear region.
Our spectra from these positions (see Figure 1k) also show
double-peaked H$\alpha$, [NII] and [S~II] emission lines.
A `loop' of radio continuum emission extends to the north from the
nuclear region (Wehrle \& Morris 1987).  These features imply the presence
of a minor-axis wind which is blowing a shell of material northward (and
perhaps southward) from the nucleus.
Thus, there is very good evidence for a minor-axis outflow in NGC~5506.

\subsubsection{IC~1368}

Our \han2 image of IC~1368 is shown in Figure 1l.
Note that the nuclear source is elongated in the direction of the
minor axis and extends $\sim$5\asec\ (1.3 kpc) on each side of the nucleus.
Such extended
emission-line halos are commonly found in edge-on starburst galaxies with
winds, so we have classified IC~1368 as a good candidate for a large-scale
outflow.

\subsubsection{IC 1417}

Our \han2 image of his galaxy shows emission from the disk, but not
from regions extending along the minor axis (Figure 1m).

\subsubsection{NGC 7590}

No extended emission is present in our \han2 image (Figure 1n).
Much of the emission comes from clumpy H~II regions in the disk.

\subsection{Individual Objects: Additional Seyferts in the Extended Sample}

For the following objects, there were no published data that is suitable for
searching for a large-scale outflow:
UGC~3255, ESO~362-G8, NGC~4117, IC~5169 and Mrk 915.

\subsubsection{NGC~513}

This galaxy has an elliptical R-band morphology (see Figure 2a).
A bright ELR of luminosity
$L_{\rm H\alpha+[NII]} =$ 1 $\times$ 10$^{39}$ erg~s$^{-1}$
is noticeable 19.7\asec\ (7.4 kpc) from the nucleus in
P.A. 210\deg.
The location of this ELR is not positioned exactly on the minor axis, but
the gas is located $\sim$4 kpc (in projection) out of the disk.  This suggests
that it has been ejected from the disk.
We have therefore classified NGC~513 as a good candidate for a large-scale
outflow.

\subsubsection{UGC~3255}

We did not obtain images of UGC~3255.  The position of the slit for our
minor-axis spectra is shown in a greyscale plot of a B-band image (digitized
survey plate) in Figure 2b.
We did not find double-peaked line profiles or
any evidence for ELRs extending out of the disk  in our minor axis spectra.

\subsubsection{ESO~362-G8}

Conical ELRs can be seen in the \han2 image of this galaxy in Figure 2c.
The northeast cone extents from the nucleus out
to $\sim$14\asec\ (4.3 kpc) along P.A. 65\deg\
(roughly perpendicular to the major axis)
and has an \han2 luminosity of 1.8 $\times$ 10$^{40}$ erg~s$^{-1}$.
Mulchaey, Wilson \& Tsvetanov (1996) find the gas in the conical ELR
to have high [O~III]5007/H$\alpha$ ratios, consistent with photoionization
by the AGN.
The presence of gas so far out of the disk suggests that a large-scale outflow
is occurring in ESO~362-G8.

\subsubsection{Mrk~10}

The presence of extranuclear ELRs in Mrk~10 has been
previously noted by Schulz (1982), who found line emission extending westward
from the nucleus.  We also find extranuclear line emission to be present
from our long-slit data.
The orientation of the slit (P.A. 40\deg) and a graph of the relative
H$\alpha$ flux
for positions along the slit is shown in Figure 2d.
Emission-line gas is present along the minor axis out to 20\asec\
(11.3 kpc) to the northeast from the nuclear region, and out to
10\asec (5.6 kpc) to the southwest.
Such emission could be produced by
a galactic outflow blowing along the minor axis.  However, due to
the limited spatial coverage of our long-slit spectra, we classify this
evidence as only suggestive.
A more complete study of the extranuclear optical emission-line gas in
Mrk~10 is clearly warranted.

\subsubsection{NGC~4117}

Our \han2 image (Figure 2e) of this galaxy shows emission from the
inner disk
but no extended emission along the minor axis.
We did not find double-peaked line profiles or
any evidence for ELRs extending out of the disk  in our minor axis spectra.

\subsubsection{NGC~5033}

A patchy ring of H~II regions shows up well in our \han2 image (Figure 2f).
The field of view of our CCD is small (5\amin) compared to the large optical
size of the galaxy ($\sim$10\amin\ $\times$ 5\amin), so our \han2 image is
not very useful for looking for emission from extraplanar gas.
The morphology of the \han2 emission within the
stellar (R-band) envelope is not elongated along the minor axis.
We did not find double-peaked line profiles or
any evidence for ELRs extending out of the disk  in our minor axis spectra.
This galaxy has been very well studied, but we did not find any evidence
in the literature for a minor axis outflow.

\subsubsection{IC~5169}

The \han2 image (Figure 2g)
shows ELRs extending along the minor axis, $\sim$6.5\asec\ (1.3 kpc) to the
northwest and southeast of the nucleus, perpendicular to the bar.
The total \han2 luminosity from the northwest ELRs is
$\sim$10$^{39}$ erg~s$^{-1}$.  The small, bright
ELR south of the nucleus has a H$\alpha$+[NII] luminosity of
$\sim$5 $\times$ 10$^{38}$ erg~s$^{-1}$ and the remaining southwest ELRs emit
$\sim$10$^{39}$ erg~s$^{-1}$.  Such emission could be produced by an outflow
along the minor axis.  However, the projected locations of the the ELRs are
inside the stellar (R-band)
envelope, so the ELRs may be located in the inner disk of IC~5169.
For this reason, we classify this evidence as only suggestive of a minor-axis
outflow.

\subsubsection{Mrk~915 (MCG~$-$2-57-23)}

No emission is evident extending along the minor axis in our \han2 image
(Figure 2h).

\section{Discussion}

\subsection{Frequency of Occurrence of Minor-Axis Outflows in Seyfert Galaxies}

The results from our observational data are listed symbolically in the last
column of Table 4.  For most of the objects we classified
as good candidates for minor-axis outflows, either
double-peaked emission lines are found from regions along the minor axis, or
the morphology of the \han2 emission resembles that of a halo extending
above and below the galaxy disk.
In NGC~513 and ESO~362-G8, the nebulae have conical morphologies,
similar to `ionization cones' observed in some Seyfert galaxies (e.g.
Pogge 1989, Wilson \& Tsvetanov 1994).
The line emission from the extranuclear gas in these objects could be
produced by ionizing radiation from the AGN or by a shock from an outflowing
jet.  However, in both cases, the ELRs are found $\sim$4 kpc (in projection)
out of the disk, which is, in general, further
out than the typical maximum extent of ionization cones in Seyferts
($\sim$2 kpc, Wilson \& Tsvetanov 1994).  An obvious possible explanation for
how the gas got so far out of the disk is that it was
blown out by a minor-axis outflow.  Thus we consider the six galaxies
NGC~2992, IC~4329A, NGC~5506, IC~1368, NGC~513 and ESO~362-G8 to be good
candidates for having large-scale minor axis outflows.

Our images and spectra of Ark~79, Mrk~10 and IC~5169 are suggestive of a
minor axis wind, but are not entirely convincing in and of themselves.
Therefore, we have not included them as `good candidates' when calculating
statistics for our samples.  Further optical studies of the minor axis nebulae
in these galaxies would be useful for determining if galactic outflows are
present.

If we ignore the results in the literature and only consider results from
the images and spectra we obtained, we find that for the 14 objects in our
representative sample, four (29\%) show good evidence for minor axis outflows.
Including the objects from Table 2, six (27\%) of the 22 objects in our
extended sample show good evidence for minor axis outflows.

As mentioned in section 4.1, there is evidence in the literature for
minor-axis outflows in two additional Seyferts in our complete sample
(NGC~4388 and NGC~4945).  All objects which show good evidence for minor
axis outflows (from our images and spectra or from the literature)
are marked with an asterisk in Tables 1 and 2.

If we include the results from the literature, we
find good evidence for minor axis outflows in
four (29\%) of the 14 objects in our representative sample,
six (27\%) of the 22 objects in our complete sample,
and
six (27\%) of the 22 objects in our extended sample.  The results are
consistent: $\sim$25$-$30\% of edge-on Seyferts
show good evidence for large-scale galactic outflows.

These results underestimate the ``true'' fraction of Seyferts which have
minor axis outflows.  For many of the objects, data suitable for detecting
emission from these outflows (e.g. H$\alpha$ images, long-slit spectra of
the minor-axis ELRs, deep radio-continuum images, deep X-ray images) are
not available.
In addition, our images are only sensitive to surface brightnesses
$\gapprox$10$^{-16}$ erg~s$^{-1}$~cm$^{-2}$~arcsec$^{-2}$
and we have
minor axis spectra to search for double-peaked emission lines for only
nine of 22 objects observed.
Detailed kinematic studies of extended emission-line regions
in Seyferts are very useful for determining if an outflow is present,
but such studies have been published for very few objects in our samples.
Assuming our edge-on samples are unbiased
subsamples of Seyfert galaxies, our results suggest that large-scale galactic
outflows are likely to be present in $\gapprox$${{1}\over{4}}$
of {\it all} Seyfert galaxies.

In Table 5, we list all Seyfert galaxies known (by us) to have good
evidence for galactic outflows.
Evidence comes in the form of optical emission-line
nebulae and split emission lines along the minor axis, extended radio continuum
emission along the minor axis, and extended X-ray halos along the minor axis.
Most of the outflows which have been identified are in
Seyferts with edge-on disks.  Outflows in only a few of these galaxies
(e.g. NGC~3079, Mrk~231 and NGC~4945) have been studied in detail.

\subsection{Energy Source for the Large-Scale Outflows}

The question of what powers large-scale outflows in Seyferts remains open
even for the most well-studied cases.
For example, in a very complete kinematic Fabry-Perot study of the two
bi-symmetric wind-blown superbubbles in the nucleus of NGC~3079,
Veilleux \etal (1994) found that either a starburst- or an AGN-driven
wind was consistent.
In the following subsections, we discuss two scenarios for powering the
large-scale outflows in Seyferts:  starburst-driven superwinds and outflows
from the active nucleus.

\subsubsection{Nuclear Starburst?}

Superwinds in starburst galaxies have been fairly well studied
(Heckman, Armus \& Miley 1990; Lehnert \& Heckman 1995), so it is natural
to compare properties of large-scale outflows in Seyferts with those of
superwinds in starburst galaxies.
The superwind in the archetypical edge-on starburst galaxy M82 has an
\han2 luminosity of $\sim$10$^{40.3}$ erg~s$^{-1}$ and a kinetic luminosity
of 10$^{42.3}$ erg~s$^{-1}$ (Heckman, Armus \& Miley 1990).
We can estimate  kinetic luminosities for the Seyfert outflows by scaling
L$^{MINOR}_{H\alpha+[NII]}$ for our sample Seyferts (Table 4) by
the ratio of the kinetic luminosity of M82 to the \han2 luminosity of
its minor-axis ELRs.
Using this method, we
find kinetic luminosities $\sim$10$^{40.5}$$-$10$^{42.2}$ erg~s$^{-1}$
(logarithmic mean 10$^{41.4}$ erg~s$^{-1}$)
for the Seyfert outflows, assuming all of the line emission is produced
by the outflow.
The implied kinetic luminosities
of our Seyfert outflows are, in general, a factor $\sim$0.1 as large as
those of superwinds from Heckman, Armus \& Miley
(1990; several $\times$ 10$^{42}$ erg~s$^{-1}$).
However, the mean far-IR luminosity of
our sample ($<\log L_{FIR}> =$ 43.4 $\log$[erg~s$^{-1}$]) is
smaller than that of Heckman, Armus \& Miley's sample
($<\log L_{FIR}> =$ 44.9 $\log$[erg~s$^{-1}$]) by about the same factor,
so if the Seyfert outflows are powered by nuclear starbursts,
smaller kinetic luminosities are to be expected.

High rates of massive star formation have been inferred for
the nuclear regions of several Seyfert galaxies.  For example, in NGC~1068,
approximately half of the IR luminosity comes from the AGN and the other half
comes from the starburst (Balick \& Heckman 1985).
The energy input from a possible starburst is directly
proportional to the starburst component of L$_{FIR}$
(cf. Heckman, Lehnert \& Armus 1993).
However, for most Seyferts, separating the starburst and AGN components of
L$_{FIR}$ is not straightforward (cf. Telesco 1988; however, see
also Rodriguez-Espinosa, Rudy \& Jones 1987, who claim that star formation
produces the bulk of the far-IR emission in Seyferts).
Measuring massive-star formation rates from stellar absorption lines is
in general, quite difficult to do in Seyferts (cf. Diaz 1992).
Thus, it is not known how common circumnuclear starbursts are in Seyferts.
Baum \etal (1993) could not distinguish between starburst- or AGN-driven winds
using their radio data.  However, they found that the luminosity of the
extranuclear radio emission is comparable to that in starburst galaxies and
follows the same radio-IR relation as that of starburst galaxies, suggesting
that the large-scale radio emission may be of starburst origin.   In order
to verify that these galactic outflows are starburst-driven superwinds, one
must find evidence for massive
stars in the Seyfert nuclei and determine if the putative starburst is
powerful enough to drive a galactic wind.

\subsubsection{Active Nucleus?}

The AGN can certainly provide enough energy to drive a galactic outflow.
However, linear nuclear radio sources (suggestive of a collimated nuclear
outflow) in Seyferts are not, in general, oriented along the same position
angle as the
large-scale diffuse radio emission (Baum \etal 1993).  Therefore, it is
unlikely that the large-scale
minor-axis radio emission is from lobes at the end of
collimated radio jets, as in powerful radio galaxies (unless the jet outflow
axis precesses).  On the other hand, if nuclear outflows from the AGN are
weak, they may lose energy to the interstellar medium in the nuclear region
and become poorly-collimated.  If the gas continues to flow outward, it
may be be diverted toward the minor axis (where the pressure gradient
is lowest) and continue on as a wide-angled outflow.

Many different models have been presented for nuclear outflows from AGNs.
Theoretical models have been proposed for hydro-magnetic jets which originate
from the accretion disk surrounding the black hole, or even from the black
hole itself (cf. Blandford 1993).  Another model
(Krolik \& Begelman 1986; Balsara \& Krolik 1993) proposes that
radiation-driven winds originate at the inner edge of the proposed
molecular torus.  Thus, estimating the amount of kinetic energy
being channelled from the AGN into the outflow is highly model-dependent.

\section{Summary and Conclusions}

Large-scale galactic outflows are quite common in Seyfert galaxies.  Our
results show that they are present in $\gapprox$${{1}\over{4}}$ of all
Seyfert galaxies.

Current observational studies of these outflows do not offer enough information
to conclude what drives the outflows.  They may be powered by
circumnuclear starbursts, AGN, or perhaps by both, in combination.  In order
to constrain the different models for the outflows, more
observational work is needed to measure physical properties
of these winds (e.g., outflow velocities, morphologies and luminosities of
radio and X-ray halos, velocity and ionization structures of the outflowing
gas near the nuclear region).

\acknowledgments

E.J.M.C. thanks Sylvain Veilleux and the referee, R.-J. Dettmar, for providing
helpful comments on the manuscript.   E.J.M C. acknowledges funding from
the Director's Office of STScI.
This research has made extensive use of the NASA/IPAC
Extragalactic Database (NED), which is operated by the Jet Propulsion
Laboratory, Caltech, under contract with NASA.
The work of M.D.L. at IGPP/LLNL was performed under the auspices of the
U.S. Department of Energy under contract W-7405-ENG-48.

\begin{planotable} {rllllllrcll}
\scriptsize
\tablecaption{ {\bf Complete Statistical Sample of Edge-on Seyfert galaxies}}
\tablehead{
\colhead{} & \colhead{Galaxy} & \colhead{Other} &
\colhead{Galaxy\tablenotemark{a}} &
\colhead{R.A.\tablenotemark{b}} & \colhead{Dec\tablenotemark{b}} &
\colhead{$\log R_{25}$\tablenotemark{b}} &
\colhead{$cz$\tablenotemark{b}} &
\multicolumn{2}{c}{Seyfert\tablenotemark{c}~~} &
\colhead{Data?} \\
\colhead{} & \colhead{Name} & \colhead{Name} &
\colhead{Type} &
\multicolumn{2}{c}{(J2000.0)}
 & \colhead{} &
\colhead{(km~s$^{-1}$)} &
\colhead{Type} & \colhead{Refs} & \colhead{}  \\
} 
\startdata
 & IC~1657      &      & (R$^\prime$)SB(s)bc & 01 14 07.3 & $-$32 39 02 &
  0.63 & 3552 & 2 & 1 & \\
 & UM~319       & MCG~0-4-112  & SB?         & 01 23 20.9 & $-$01 58 36 &
  0.43 & 4730 & 2 & 1,2,3 & \\
 & Mrk~993      &      UGC~987 & Sa          & 01 25 31.5 & $+$32 08 10 &
  0.52 & 4658 & 2 & 1,3 &\dag\\
 & Mrk~577      &              & S0/a        & 01 49 30.1 & $+$12 30 32 &
  0.41 & 5179 & 2 & 3 & \\
 & Ark~79       & UGC~1757     & S?          & 02 17 23.1 & $+$38 24 50 &
  0.53 & 5157 & 2 & 2,3 &\dag\\
 & NGC~931      & Mrk~1040     & Sbc         & 02 28 15.0 & $+$31 18 46 &
  0.67 & 4993 & 1 & 1,2,3 &\dag\\
 & NGC~1320     & Mrk~607      & Sa: sp      & 03 24 49.0 & $-$03 02 31 &
  0.47 & 2700 & 2 & 2,3 &\dag\\
 & NGC~1386     &              & SB(s)0+     & 03 36 46.4 & $-$35 59 58 &
  0.42 &  864 & 2 & 1,2,3 &\dag\\
{*}& NGC~2992   &              & Sa pec      & 09 45 42.1 & $-$14 19 39 &
  0.51 & 2314 & 2 & 1,3 &\dag\\
 & MCG~$-$2-27-9 &           & SB(rs)0+ pec? & 10 35 27.4 & $-$14 07 49 &
  0.60 & 4650 & 2 & 3 &\dag\\
 & NGC~4235     &      IC~3098 & SA(s)a      & 12 17 08.9 & $+$07 11 31 &
  0.65 & 2410 & 1 & 1,2,3 &\dag\\
{*}& NGC~4388   &              & SA(s)b: sp  & 12 25 47.0 & $+$12 39 42 &
  0.64 & 2517 & 2 & 1 & \\
 & NGC~4602     &              & SAB(rs)bc   & 12 40 37.1 & $-$05 07 58 &
  0.46 & 2548 & 1.9 & 3 &\dag\\
{*}& NGC~4945   &              & SB(s)cd: sp & 13 05 26.2 & $-$49 28 15 &
  0.72 &  560 & 2 & 3 & \\
{*}& IC~4329A   &              & SA0+: sp    & 13 49 19.4 & $-$30 18 35 &
  0.56 & 4793 & 1 & 1,2,3 &\dag\\
{*}& NGC~5506   &              & Sa pec sp   & 14 13 14.9 & $-$03 12 27 &
  0.52 & 1815 & 2 & 1,2,3 &\dag\\
 & ESO~103-G35  &              & SA0         & 18 38 20.3 & $-$65 25 42 &
  0.44 & 3983 & 2 & 1,3 & \\
 & NGC~6810     &              & Sa          & 19 43 34.2 & $-$58 39 21 &
  0.55 & 1958 & 2 & 3 & \\
{*}& IC~1368    &              & Sa? sp      & 21 14 12.1 & $+$02 10 38 &
  0.44 & 3912 & 2 & 3 &\dag\\
 & IC~1417      &              & Sb? sp      & 22 00 21.6 & $-$13 08 48 &
  0.58 & 4309 & 2 & 1,4 &\dag\\
 & NGC~7410     &              & SB(s)a      & 22 55 00.7 & $-$39 39 41 &
  0.51 & 1751 & 2 & 2,3 & \\
{ }& NGC~7590   &              & S(r?)bc     & 23 18 55.0 & $-$42 14 17 &
  0.42 & 1596 & 2 & 2,3 &\dag\\
\tablenotetext{*}{Evidence exists for a minor-axis outflow (see section 4)}
\tablenotetext{a}{Host galaxy morphology, taken from the NASA Extragalactic
  Database (NED)}
\tablenotetext{b}{R.A., declination and Axial ratios
  ($R_{25}$) were taken from RC3.
  Recessional velocities ($cz$) were taken from (in preferential
  order): 21~cm values
  listed in RC3, optical values listed in RC3, Huchra 1993, Veron-Cetty
  \& Veron 1991, and Hewitt \& Burbidge 1991.}
\tablenotetext{c}{Seyfert type.  References: (1) Huchra 1993,
  (2) Veron-Cetty \& Veron 1991, (3) Hewitt \& Burbidge 1991, and
  (4) Maia et al. 1987}
\tablenotetext{\dag}{Galaxy is in our representative sample -- new
optical data are presented for it in the present paper.  }
\end{planotable}

\begin{planotable}{rllllllrcl}
\scriptsize
\tablecaption{ {\bf Additional Edge-on Seyferts in Extended Sample}}
\tablehead{
\colhead{} & \colhead{Galaxy} & \colhead{Other} &
\colhead{Galaxy\tablenotemark{a}} &
\colhead{R.A.\tablenotemark{b}} & \colhead{Dec\tablenotemark{b}} &
\colhead{$\log R_{25}$\tablenotemark{b}} &
\colhead{$cz$\tablenotemark{b}} &
\multicolumn{2}{c}{Seyfert\tablenotemark{c}~~} \\
\colhead{} & \colhead{Name} & \colhead{Name} &
\colhead{Type} &
\multicolumn{2}{c}{(J2000.0)} &
\colhead{} &
\colhead{(km~s$^{-1}$)} &
\colhead{Type} & \colhead{Refs} \\
} 
\startdata
{*}& NGC~513    & Ark~41         & S?          & 01 24 27.0 & $+$33 47 57 &
  0.31 & 5949 & 2 & 1,2,3 \\
 & UGC~3255     &                & SBb?        & 05 09 48.0 & $+$07 29 00 &
  0.63 & 5689 & 2 & 1,2,3 \\
{*}& ESO~362-G8 &                & S0?         & 05 11 09.1 & $-$34 23 36 &
  0.34 & 4785 & 2 & 2,3 \\
 & Mrk~10       &                & SBbc        & 07 47 29.1 & $+$60 55 59 &
  0.38 & 8753 & 1 & 1,2,3 \\
 & NGC~4117     &                & S0:         & 12 07 46.2 & $+$43 07 34 &
  0.31 &  871 & 2 & 1,2 \\
 & NGC~5033     &                & SA(s)c      & 13 13 28.0 & $+$36 35 38 &
  0.33 &  878 & 1 & 1,2 \\
 & IC~5169      &            & (R$_1$)SAB(r)0+ & 22 10 09.7 & $-$36 05 20 &
  0.37 & 3016 & 2 & 1,4 \\
 & Mrk~915      & MCG~$-$2-57-23 & S?          & 22 36 46.3 & $-$12 32 40 &
  0.45 & 7248 & 1 & 1,2 \\
\tablenotetext{}{ Edge-on Seyferts which we observed which did not satisfy the
  selection criteria for the complete sample (see section 2) }
\tablenotetext{*}{Evidence exists for a minor-axis outflow (see section 4).}
\tablenotetext{a}{Host galaxy morphology, taken from the NASA Extragalactic
  Database (NED)}
\tablenotetext{b}{R.A. and declination and Axial ratios ($R_{25}$) were taken
  from RC3.  Recessional
  velocities ($cz$) are optical (or 21~cm, when available) velocities listed
  in RC3.}
\tablenotetext{c}{References for Seyfert types: (1) Huchra 1993,
  (2) Veron-Cetty \& Veron 1991, (3) Hewitt \& Burbidge 1991, and
  (4) Maia et al. 1987}
\end{planotable}

\begin{planotable}{llccc}
\scriptsize
\tablecaption{ {\bf Observing Log}}
\tablehead{
\colhead{Date} &
\colhead{Type} & \colhead{Instruments} &
\colhead{Pixel} & \colhead{Wavelength Coverage} \\
\colhead{} &
\colhead{} & \colhead{} & \colhead{Size} & \colhead{} \\
} 
\startdata
Jan 92 & Long-slit Spectroscopy & GoldCam Spectrometer &
0.78\asec\ $\times$ 1.24\AA  & 6200 $-$ 7200 \AA \\
       &                        & TI 800 $\times$ 800 CCD & & \\
Jun 92 & Long-slit Spectroscopy & GoldCam Spectrometer &
0.78\asec\ $\times$ 1.24\AA  & 4800 $-$ 7100 \AA \\
       &                        & Ford 3k $\times$ 1k CCD & & \\
Aug 92 & Direct Imaging & Tek 1k $\times$ 1k CCD &
0.3\asec\ & R,H$\alpha$ \\
Feb 93 & Direct Imaging & Tek 1k $\times$ 1k CCD &
0.3\asec\ & R,H$\alpha$ \\
\tablenotetext{}{ All observations were performed at the 2.1~m telescope at
  KPNO. }
\end{planotable}

\begin{planotable}{lcccccllll}
\scriptsize
\tablecaption{ {\bf Data for Galaxies in Extended Sample} }
\tablehead{
\colhead{Galaxy} &
\colhead{Images\tablenotemark{1}} &
\colhead{Source} & \colhead{Long Slit Spectra} &
\colhead{Source} &
\colhead{Fig} &
L$^{\rm TOTAL}_{H\alpha}$$^{,3}$ &
L$^{\rm MINOR}_{H\alpha}$$^{,3}$ &
L$_{FIR}$$^{~4}$ &
\colhead{Evidence\tablenotemark{5}} \\
\colhead{Name} &
\colhead{} & \colhead{of Data\tablenotemark{2}} & \colhead{axis (P.A.)} &
\colhead{of Data\tablenotemark{2}} &
\colhead{no} &
\colhead{} \\
} 
\startdata
Mrk~993  & R,H$\alpha$+[NII] &1& maj (32\deg), min (122\deg) &2& 1a
&40.87&{...}&43.08 &   \\
Ark~79   & R,H$\alpha$+[NII] &1& maj (87\deg), min (177\deg) &2& 1b
&40.95&{...}&{...} & S? \\
NGC~931  & R                 &1& maj (73\deg), min (163\deg) &2& 1c
&{...}&{...}&43.87 &   \\
NGC~1320 & R,H$\alpha$+[NII] &1&                    {...}    &{...}& 1d
&40.62&{...}&43.21 &   \\
NGC~1386 & 6939\AA, H$\alpha$+[NII] &3&    {...}        &{...}& 1e
&40.31&{...}&43.15  &   \\
NGC~2992 & 6680\AA, H$\alpha$+[NII] &3&    {...}        &{...}& 1f
&41.11&38.5&43.66 & I \\
MCG~$-$2-27-9 & R,H$\alpha$+[NII] &4&           {...}    &{...}& 1g
&40.94&{...}&43.21 &   \\
NGC~4235 & R,H$\alpha$+[NII] &5& maj (48\deg), min (138\deg) &2&1h
&40.38&{...}&42.36 & \\
NGC~4602 & 5265\AA, H$\alpha$+[NII] &3&    {...}        &{...}& 1i
&41.32&{...}&43.64 &  \\
IC~4329A & R,H$\alpha$+[NII] &4&                {...} &{...}& 1j
&41.46&39.4&43.63 & I \\
NGC~5506 & 5265\AA, H$\alpha$+[NII] &3& maj (91\deg), min(1\deg) &2& 1k
&40.74&39.7&43.43 & I,S \\
IC~1368  & R,H$\alpha$+[NII] &1& maj (45\deg)                &6& 1l
&40.56&39.5&43.82 & I \\
IC~1417  & R,H$\alpha$+[NII] &1&                    {...}    &{...}& 1m
&40.78&{...}&43.14 & \\
NGC~7590 & 5159\AA, H$\alpha$+[NII]&3&         {...}    &{...}& 1n
&41.29&{...}&43.40 & \\
\\
\tableline
\\
NGC~513  & R, H$\alpha$+[NII] &1&                    {...}    &{...}& 2a
&41.42&39.0&43.93 & I \\
UGC~3255 &     {...}         &{...}& maj (17\deg), min (107\deg) &2& 2b
&{...}&{...}&43.68 &  \\
ESO~362-G8 & R, H$\alpha$+[NII] &4&              {...}        &{...}& 2c
&41.07&40.2&43.18 & I \\
Mrk~10   &     {...}         &{...}& maj (130\deg), min (40\deg) &2& 2d
&{...}&{...}&43.92 & S? \\
NGC~4117 & 6520\AA, H$\alpha$+[NII] &4& maj (18\deg), min (108\deg) &2& 2e
&39.38&{...}&{...} & \\
NGC~5033 & R, H$\alpha$+[NII]$^6$ &5& maj (170\deg), min (80\deg) &2& 2f
&40.68$^6$&{...}&43.22 & \\
IC~5169  & R, H$\alpha$+[NII] &4&                    {...}    &{...}& 2g
&40.57&{...}&43.58 & I? \\
Mrk~915  & R, H$\alpha$+[NII] &1&                    {...}    &{...}& 2h
&41.56&{...}&43.49 & \\
\tablenotetext{}{The first 14 objects are those from our complete sample
  (i.e., this is our representative sample).  All of the objects listed
  make up our extended sample.}
\tablenotetext{1}{Images listed by Angstrom are narrow-band
  (FWHM $\sim$50$-$100\AA) continuum images centered at that wavelength.}
\tablenotetext{2}{Sources of Data: (1) images from the Aug 92 run, (2) spectra
  from the Jan 92 run, (3) images from Tsvetanov, Fosbury \& Tadhunter (1995),
  (4) images from Mulchaey, Wilson \& Tsvetanov (1996), (5) images from the
  Feb 93 run, and (6) spectra from the Jun 92 run.}
\tablenotetext{3}{Total luminosity \han2 luminosity
  (L$^{\rm TOTAL}_{H\alpha}$) and \han2 luminosity of minor axis
  emission-line regions (L$^{\rm MINOR}_{H\alpha}$; see section 4 for
  details).  Units are log(erg~s$^{-1}$).  L$^{\rm TOTAL}_{H\alpha}$
  is accurate to $\sim$20\% ($\sim$30\% for NGC~4235 and IC~5033) and
  L$^{\rm MINOR}_{H\alpha}$ is accurate to $\sim$50\%.}
\tablenotetext{4}{Far infrared luminosities in units of log(erg~s$^{-1}$),
  calculated using the method described in Fullmer \& Lonsdale (1989).
  IRAS fluxes are
  from Moshir \etal (1990) (taken from the NASA Extragalactic Database [NED]),
  except for UGC~3255, ESO~362-G8, NGC~2992 and NGC~7590.  IRAS fluxes for
  these galaxies were taken from the IRAS Point Source Catalog (1988).  }
\tablenotetext{5}{Evidence for minor axis wind from images (I) or long-slit
  spectral data (S).  Question mark indicates only suggestive evidence.}
\tablenotetext{6}{CCD field of view is smaller than the optical size of the
  galaxy.  The luminosity listed is for the field of view covered by the CCD.}
\end{planotable}

\begin{planotable}{lcrlll}
\scriptsize
\tablecaption{ {\bf Minor-Axis Outflows in Seyferts}}
\tablehead{
\colhead{Galaxy} &  \colhead{Seyfert} &
\colhead{$cz$\tablenotemark{1}} & \colhead{$\log R_{25}$\tablenotemark{1}} &
\multicolumn{2}{l}{~~~~~Evidence for Minor-axis Outflows} \\
\colhead{Name} & \colhead{Type} &
\colhead{(km~s$^{-1}$)} & \colhead{} &
\multicolumn{1}{l}{References\tablenotemark{2}} &
\multicolumn{1}{l}{Notes\tablenotemark{3}} \\
} 
\startdata
NGC~513  & 2 & 5949 & 0.31 & OI (1) &
 bright ELR approx along min axis \\
&&&&& possibly ion cone \\
ESO~362-G8 & 2 & 4830 & 0.34 & OI (1) &
 v bright large-scale conical ELRs on min axis \\
&&&&& possibly ionized by AGN \\
NGC~2992 & 2  & 2314 & 0.51 & OI (1),RI (2) &
 many bright large-scale ELRs \\
&&&&& radio bubble, v good evidence \\
NGC~3079 & 2?\tablenotemark{4} & 1125 & 0.74 & OI,OS,RI,XI (3,4,5) &
 definite wind, well studied \\
&&&&& could be starburst- or AGN-driven \\
NGC~4388 &  2 & 2517 & 0.64 & OI,OS,RI (6,7,8) &
 radial outflow of optically emitting gas \\
&&&&& radio finger perp to disk, good evidence \\
Mrk~231  & 1 & 12300 & 0.13 & OS (9), RI (10)  &
 definite wind \\
&&&&& could be starburst- or AGN-driven \\
NGC~4945 & 2 & 560 & 0.72 & OI,OS (3,4) &
 definite wind, well studied \\
&&&&& could be starburst- or AGN-driven \\
IC~4329A & 1 & 4800 & 0.56 & OI (1) &
 bi-symmetric EL halo on min axis \\
&&&&& \\
NGC~5506 & 2  & 1830 & 0.52 & OI,OS,RI (1,11,12) &
 double-peaked ELs  from min axis ELRs \\
&&&&& radio bubble, v good evidence \\
NGC~5548 & 1 & 5149 & 0.05 & OI,OS (13), RI (10) &
 definite min axis outflow\\
&&&&& may be from AGN\\
NGC~5728 & 2 & 2790 & 0.24 & OI,OS (14) &
 conical ELRs and outflow along min axis\\
&&&&& may be from AGN \\
IC~1368  & 2 & 3912 & 0.44 & OI (1) &
 bi-symmetric EL halo along min axis \\
\tablenotetext{1}{Recessional velocities ($cz$) and axial ratios ($R_{25}$)
   from RC3.  If available, 21~cm velocities are quoted.}
\tablenotetext{2}{References for evidence for minor-axis outflows
  (OI = optical images, OS = optical spectra, RI = radio continuum images,
  XI = X-ray images):
   (1) this paper,
   (2) Wehrle \& Morris 1988, (3) Heckman, Armus \& Miley 1990,
   (4) Heckman, Lehnert \& Armus 1993,
   (5) Veilleux \etal 1994, (6) Corbin
   \etal 1988, (7) Hummel \etal 1983, (8) Stone \etal 1988,
   (9) Hamilton \& Keel 1987, (10) Baum \etal 1993,
   (11) Wilson, Baldwin \& Ulvestad 1985, (12) Wehrle \& Morris
   1987, (13) Wilson \etal 1989, (14) Schommer \etal 1988.
   }
\tablenotetext{3}{Abbreviations for notes: EL = emission-line, ELR =
   emission-line region, ion = ionization, min = minor, poss = possible,
   v = very, perp = perpendicular}
\tablenotetext{4}{NGC~3079 has a LINER-type optical nuclear spectrum, but
  it may house a weak Seyfert nucleus.}
\end{planotable}

\clearpage

\noindent
{\bf Figure 1 -- }
Images and minor-axis spectra for the 14 objects observed in
our complete sample.  Levels for all of the contour plots of the images differ
by a multiplicative factor of 2.  The highest level for the contour plots of
the R-band and continuum images is 50\% of the maximum surface brightness.
Slit positions for the minor-axis spectra are overlayed on the contour plots.
{\bf (a) Mrk~993. }
   Contour plots of the R-band emission (left) and the \han2
   emission (right).
   The lowest level displayed in the \han2 contour plot is
   0.5 $\times$ 10$^{-16}$ erg~s$^{-1}$~cm$^{-2}$~arcsec$^{-2}$.
{\bf (b) Ark~79.}
   Contour plots of the R-band emission (left) and the \han2
   emission (right).
   The lowest level displayed in the \han2 contour plot is
   0.5 $\times$ 10$^{-16}$ erg~s$^{-1}$~cm$^{-2}$~arcsec$^{-2}$.
   Also shown are spectra (effective aperture 2.0\asec\ $\times$ 2.34\asec)
   from the nucleus and from regions 3.1\asec\ (1 kpc)
   north and south of the nucleus, along the minor axis.
{\bf (c) NGC~931.}
   Contour plot of the R-band emission (left).  Emission was detected
   along the minor axis, but it is from the companion galaxy
   located $\sim$20\asec\ to the north.
{\bf (d) NGC~1320.}  Its companion galaxy NGC~1321 lies $\sim$1.7\amin\
   to the north of NGC~1320.
   Contour plots of the R-band emission (left) and the \han2
   emission (right).
   The lowest level displayed in the \han2 contour plot is
   1.0 $\times$ 10$^{-16}$ erg~s$^{-1}$~cm$^{-2}$~arcsec$^{-2}$.
{\bf (e) NGC~1386.}
   Contour plots of the red continuum emission (left) and
   the \han2 emission (right).
   The lowest level displayed in the \han2 contour plot is
   2.0 $\times$ 10$^{-16}$ erg~s$^{-1}$~cm$^{-2}$~arcsec$^{-2}$.
{\bf (f) NGC~2992.}
   Contour plots of the red continuum emission (left) and
   the \han2 emission (right).
   The lowest level displayed in the \han2 contour plot is
   1.0 $\times$ 10$^{-16}$ erg~s$^{-1}$~cm$^{-2}$~arcsec$^{-2}$.
{\bf (g) MCG~$-$2-27-9.}
   Contour plots of the R-band emission (left) and the \han2
   emission (right).
   The lowest level displayed in the \han2 contour plot is
   2.0 $\times$ 10$^{-16}$ erg~s$^{-1}$~cm$^{-2}$~arcsec$^{-2}$.
{\bf (h) NGC~4235.}
   Contour plots of the R-band emission (left)
   the \han2 emission (right).
   The lowest level displayed in the \han2 contour plot is
   1.0 $\times$ 10$^{-16}$ erg~s$^{-1}$~cm$^{-2}$~arcsec$^{-2}$.
{\bf (i) NGC~4602.}
   Contour plots of the green continuum emission (left) and the \han2
   emission (right).
   The lowest level displayed in the \han2 contour plot is
   2.0 $\times$ 10$^{-16}$ erg~s$^{-1}$~cm$^{-2}$~arcsec$^{-2}$.
{\bf (j) IC~4329A.}
   Contour plots of the R-band emission (left) and the \han2
   emission (right).
   The lowest level displayed in the \han2 contour plot is
   2.0 $\times$ 10$^{-16}$ erg~s$^{-1}$~cm$^{-2}$~arcsec$^{-2}$.
{\bf (k) NGC~5506.}
   Contour plots of the green continuum emission (left)
   and the \han2 emission (right).
   The lowest level displayed in the \han2 contour plot is
   1.0 $\times$ 10$^{-15}$ erg~s$^{-1}$~cm$^{-2}$~arcsec$^{-2}$.
   Also shown are spectra (effective aperture 2.0\asec\ $\times$ 2.34\asec)
   from the nucleus and from regions 6.2\asec\ (0.7 kpc)
   north and south of the nucleus, along the minor axis.
{\bf (l) IC~1368.}
   Contour plots of the R-band emission (left)
   and the \han2 emission (right).
   The lowest level displayed in the \han2 contour plot is
   1.0 $\times$ 10$^{-16}$ erg~s$^{-1}$~cm$^{-2}$~arcsec$^{-2}$.
{\bf (m) IC~1417.}
   Contour plots of the R-band emission (left)
   and the \han2 emission (right).
   The lowest level displayed in the \han2 contour plot is
   0.5 $\times$ 10$^{-16}$ erg~s$^{-1}$~cm$^{-2}$~arcsec$^{-2}$.
{\bf (n) NGC~7590.}
   Contour plots of the green continuum emission (left)
   and the \han2 emission (right).
   The lowest level displayed in the \han2 contour plot is
   4.0 $\times$ 10$^{-16}$ erg~s$^{-1}$~cm$^{-2}$~arcsec$^{-2}$.

\clearpage

\noindent
{\bf Figure 2 -- }
Images and minor-axis spectra for eight objects observed which were not in
our complete sample.
Levels for all of the contour plots of the images differ
by a multiplicative factor of 2.  The highest level for the contour plots of
the R-band and continuum images is 50\% of the maximum surface brightness.
Slit positions for the minor-axis spectra are overlayed on the contour plots.
{\bf (a) NGC~513. }
   Contour plots of the R-band emission (left) and the \han2 emission (right).
   The lowest level displayed in the \han2 contour plot is
   0.5 $\times$ 10$^{-16}$ erg~s$^{-1}$~cm$^{-2}$~arcsec$^{-2}$.
{\bf (b) UGC~3255.}
   Linearly-scaled greyscale plot of the B-band emission
   from a digitized sky survey plate obtained at STScI (left).
{\bf (c) ESO~362-G8.}
   Contour plots of the R-band emission (left) and the \han2
   emission (right).
   The lowest level displayed in the \han2 contour plot is
   1.0 $\times$ 10$^{-16}$ erg~s$^{-1}$~cm$^{-2}$~arcsec$^{-2}$.
{\bf (d) Mrk~10.}
   Linearly-scaled greyscale plot of the B-band emission
   from a digitized sky survey plate obtained at STScI (left).
   We also show the relative intensity of
   the H$\alpha$ line emission along the minor-axis slit.
{\bf (e) NGC~4117.}
   Contour plots of the red continuum emission (left) and the \han2
   emission (right).
   The lowest level displayed in the \han2 contour plot is
   1.0 $\times$ 10$^{-16}$ erg~s$^{-1}$~cm$^{-2}$~arcsec$^{-2}$.
{\bf (f) NGC~5033.}
   Contour plots of the R-band emission (left) and the \han2
   emission (right).
   The lowest level displayed in the \han2 contour plot is
   1.0 $\times$ 10$^{-16}$ erg~s$^{-1}$~cm$^{-2}$~arcsec$^{-2}$.
   Note that the field of view of the CCD was smaller than the extent of
   the emission from the galaxy.
{\bf (g) IC~5169.}
   Contour plots of the R-band emission (left) and the \han2
   emission (right).
   The lowest level displayed in the \han2 contour plot is
   2.0 $\times$ 10$^{-16}$ erg~s$^{-1}$~cm$^{-2}$~arcsec$^{-2}$.
{\bf (h) Mrk~915.}
   Contour plots of the R-band emission (left) and the \han2
   emission (right).
   The lowest level displayed in the \han2 contour plot is
   1.0 $\times$ 10$^{-16}$ erg~s$^{-1}$~cm$^{-2}$~arcsec$^{-2}$.


\begin{references}
\def\rf{\reference}

\rf Amram, P., Marcelin, M., Bonnarel, F., Boulesteix, J., Afanasiev, V. L.,
    \& Dodonov, S. N. 1992, A\&A, 263, 69
\rf Antonucci, R. R. J. 1993, ARA\&A, 31, 473
\rf Balick, B., \& Heckman, T. M. 1985, AJ, 90, 197
\rf Balsara, D. S., \& Krolik, J. J. 1993, \apj, 402, 109
\rf Baum, S. A., O'Dea, C. P., Dallacassa, D., de Bruyn, A. G., \&
    Pedlar, A. 1993, \apj, 419, 553
\rf Blandford, R. D. 1993, in Astrophysical Jets, ed. D. Burgarella, M. Livio
\&
    C.~P. O'Dea (Cambridge: Cambridge University Press), p.~15
\rf Bregman, J. N. 1994, in The Soft X-ray Cosmos: Proc. of the ROSAT
  Science Symposium, ed. E.~Schlegel \& R.~Petre (New York: AIP), p.~164
\rf Buta, R., \& de Vaucouleurs, G. 1983, \apjs, 51, 149
\rf Corbin, M. R., Baldwin, J. A., \& Wilson, A. S. 1988, \apj, 334, 584
\rf Dettmar, R.-J. 1992, Fund. Cosmic Phys., 15, 143
\rf Diaz, A. I. 1992, in Relationships Between Active Galactic Nuclei and
    Starburst Galaxies, ed. A.~V. Filippenko (San Francisco: ASP), p.~181
\rf de Vaucouleurs, G., de Vaucouleurs, A., Corwin, H. G., Buta, R. J.,
    Paturel, G., \& Fouque, P. 1991, Third Reference Catalog of Bright
    Galaxies (RC3; New York: Springer-Verlag)
\rf Fullmer, L., \& Lonsdale, C. 1989, Cataloged Galaxies and Quasars in the
      IRAS Survey (JPL Pub. D-1932, Version 2, Appendix B)
\rf Hamilton, D., \& Keel, W. C. 1987, \apj, 321, 211
\rf Harnett, J., Haynes, R., Klein, U., \& Wielebinski, R. 1989, A\&A, 216, 39
\rf Heckman, T. M., Armus, L., \& Miley, G. K. 1990, \apjs, 74, 833
\rf Heckman, T. M., Lehnert, M. D., \& Armus, L. 1993, in The Environment
    and Evolution of Galaxies, ed. J.~M. Shull \& H.~A. Thronson
    (Dordrecht: Kluwer), p.~455
\rf Hewitt, A. \& Burbidge, G. 1991, \apjs, 75, 297
\rf Huchra, J. 1993, private communication (updated electronic version of
     ``Catalogue of Seyfert Galaxies and Other Bright AGN'')
\rf Hummel, E., Beck, R., \& Dettmar, R.-J. 1991, A\&AS, 87, 309
\rf Hummel, E., van Gorkom, J., \& Kotanyi, C. 1983, \apj, 267, L5
\rf Keel, W. C., \& Wehrle, A. E. 1993, AJ, 106, 236
\rf Krolik, J. H., \& Begelman, M. C. 1986, \apj, 308, L55
\rf Lehnert, M. D., \& Heckman, T. M. 1995, \apjs, 97, 89
\rf Maia, M. A. G., da Costa, L. N., Willmer, C., Pellegrini, P. S. \& Rite, C.
    1987, AJ, 93, 546
\rf Moshir, et al. 1990, IRAS Faint Source Catalog, version 2
\rf Mulchaey, J. S., Wilson, A. S., \& Tsvetanov, Z. 1996, ApJS, in press
  (Feb 10 issue)
\rf Pogge, R. W. 1988, \apj, 332, 702
\rf Pogge, R. W. 1989, \apj, 345, 730
\rf Rand, R. J. 1995, in Gas Disks in Galaxies, ed. J. M. van der Hulst
    (The Hague: Kluwer), in press
\rf Rodriguez-Espinosa, J. M., Rudy, R. J., \& Jones, B. 1987, \apj, 312, 555
\rf Schommer, R. A., Caldwel, N., Wilson, A. S., Baldwin, J. A.,
    Phillips, M. M., Williams, T.~B., \& Turtle, A.~J. 1988, \apj, 324, 154
\rf Schulz, H. 1982, A\&A, 115, 209
\rf Scoville, N. Z. 1992, in Relationships Between Active Galactic Nuclei and
    Starburst Galaxies, ed. A.~V. Filippenko (San Francisco: ASP), p.~159
\rf Seaquist, E. \& Odegard, N. 1991, \apj, 369, 320
\rf Stone, J. L., Jr., Wilson, A. S., \& Ward, M. J. 1988, \apj, 330, 105
\rf Telesco, C. M. 1988, ARA\&A, 26, 343
\rf Terlevich, R. 1992, in Relationships Between Active Galactic Nuclei and
    Starburst Galaxies, ed. A.~Filippenko (San Francisco: ASP), p.~133
\rf Tsvetanov, Z., Dopita, M., \& Allen, M. 1995, B.A.A.S.,
    Vol.~26, No.~5, p.~871
\rf Tsvetanov, Z. I., Fosbury, R. A. E., \& Tadhunter, C. N. 1995,
    ApJS, submitted
\rf Ulvestad, J. S., \& Wilson, A. S. 1984, \apj, 285, 439
\rf Ulvestad, J. S., \& Wilson, A. S. 1989, \apj, 343, 659
\rf Unger, S. W., Lawrence, A., Wilson, A. S., Elvis, M., \& Wright, A. E.
    1987, MNRAS, 228, 521
\rf Veilleux, S., Cecil, G., Bland-Hawthorn, J., Tully, R. B.,
    Filippenko, A. V., \& Sargent, W.~L.~W. 1994, \apj, 433, 48
\rf Veron-Cetty, M. P., \& Veron, P. 1991, electronic version of
    ``A Catalog of Quasars and Active Nuclei,'' 5th ed. (ESO Sci. Rep. No. 10)
\rf Weaver, K. A., Wilson, A. S., \& Baldwin, J. A. 1991, \apj, 366, 50
\rf Wehrle, A. E., \& Morris, M. 1987, \apj, 313, L43
\rf Wehrle, A. E., \& Morris, M. 1988, \aj, 95, 1689
\rf Wilson, A. S. 1993, in Astrophysical Jets, ed. D. Burgarella, M. Livio \&
    C.~P. O'Dea (Cambridge: Cambridge University Press), p.~121
\rf Wilson, A. S., Baldwin, J. A., \& Ulvestad, J. S. 1985, \apj, 291, 627
\rf Wilson, A. S., \& Tsvetanov, Z. I. 1994, AJ, 107, 1227
\rf Wilson, A. S., Wu, X., Heckman, T. M., Baldwin, J. A., \& Balick, B. 1989,
     \apj, 339, 729

\end{references}
\end{document}